\pdfoutput=1
\documentclass[12pt]{article}

\usepackage{amsmath, amsthm, amsfonts, amssymb, bm, bbm}
\usepackage{algorithm, algorithmic, graphicx, subcaption, epstopdf}
\usepackage{hyperref, xcolor, tabularx, multirow, booktabs, enumerate, natbib, comment}
\usepackage[normalem]{ulem}

\def\spacingset#1{\renewcommand{\baselinestretch}%
{#1}\small\normalsize} \spacingset{1}

\newtheorem{theorem}{Theorem}
\newtheorem{assumption}{Assumption}

\newtheorem{proposition}{Proposition}

\newtheorem{definition}{Definition}

\newcommand{\cB}{\mathcal{B}}
\newcommand{\cC}{\mathcal{C}}
\newcommand{\cF}{\mathcal{F}}
\newcommand{\cH}{\mathcal{H}}

\newcommand{\cS}{\mathcal{S}}
\newcommand{\cU}{\mathcal{U}}

\DeclareMathOperator*{\argmin}{argmin}
\newcommand{\bigtimes}{\mathop{\scalebox{1.25}{$\times$}}}
\def\bfN{N}

\begin{document}

\title{\Large{\textbf{Point Process Structural Equation Models}}} 
\author{
\bigskip
\large{Shuoxun Xu$^\dag$, Kuang-Yao Lee$^\ddag$, and Lexin Li$^\dag$\thanks{Address for correspondence: Lexin Li, Department of Biostatistics and Epidemiology, University of California, Berkeley, 2121 Berkeley Way, \#5302, Berkeley, CA 94720 U.S.A. \href{mailto:lexinli@berkeley.edu}{lexinli@berkeley.edu}}} \\
\normalsize{$^\dag$\textit{University of California at Berkeley} and $^\ddag$\textit{Temple University}} \\
}
\date{}
\maketitle

\spacingset{1.2} 
\begin{abstract}
Multivariate point process data arise in numerous scientific applications, where a central goal is to recover directed relationships among multiple event processes. However, existing methods offer limited solutions for learning directed networks directly from point process data. In this article, we propose a nonparametric structural equation model for multivariate point process based on an additive Cox process framework. Our approach represents point processes as random counting measures and employs nested kernel embedding to accommodate flexible nonlinear dependencies while preserving event-time information. We develop a two-stage estimation procedure. First, we recover the topological ordering of the graph using a new variance-to-mean ratio criterion that exploits the overdispersion property of the Cox process and identifies directional relations. Second, we identify the parent sets through regularized process-on-process regression with functional regression operators. We establish theoretical guarantees for consistent ordering identification, parent selection, and graph recovery, while allowing the number of point processes to diverge with the sample size. We demonstrate improved graph recovery over a count-based alternative through numerical simulations. We further apply our method to multiregion neural spike trains, revealing directed neural associations and their variations across experimental conditions.
\end{abstract}
\bigskip


\newpage
\spacingset{1.8}

\section{Introduction}
\label{sec:introduction}

Multivariate point process data commonly arise across many scientific domains, where multiple sequences of discrete events are observed over time. Examples include neural spike trains recording action potentials from populations of neurons \citep{cai2024latent,pinkney2025identifying}, electronic health records tracking sequences of medical events \citep{sun2021recurrent,knight2025latent}, social networks capturing user interactions \citep{zipkin2016point,zhang2023learning}, and financial systems monitoring transactions and market activities \citep{xu2020semiparametric,xu2024marked}. In these applications, each process is represented as an event sequence occurring at irregular times. One of the central goals is to infer the directed relationships among these event processes, namely, to determine how events in one process trigger, influence, or propagate to events in another. Recovering such directed network structure is crucial for understanding information flow in neural circuits, disease progression in healthcare systems, and influence dynamics in social and economic networks. Despite its importance, however, there is currently no solution directly applicable to this problem.

Our motivation comes from modern neuroscience experiments that record simultaneous neural activity from populations of neurons during sensory and behavioral tasks. Specifically, \citet{steinmetz2019distributed} used Neuropixels probes to record activity from about 30,000 neurons across multiple mouse brain regions while animals performed a visual discrimination task. In this study, each neuron generates a sequence of firing events over time, commonly referred to as a spike train. The resulting data are naturally represented as collections of event times observed across multiple neurons and repeated trials. A fundamental scientific question of interest is to understand how neural activity propagates through neural circuits, i.e., which neurons exhibit directed influence on others and how such influence depends on temporal firing patterns. However, this problem is particularly challenging because the observations are discrete and irregular event processes, rather than continuous trajectories or vector-valued measurements. Moreover, neural interactions can be nonlinear, with downstream activity depending not only on firing rates but also on temporal and burst patterns of upstream spikes. These challenges motivate us to develop a structural equation modeling (SEM) framework for multivariate point process data that can infer directed relationships among event processes while accommodating flexible dependencies.

There have been several related lines of research. The first line is classical SEM, which operates on random variables by specifying how each variable depends on its parent variables together with additive noise \citep{pearl2009causality, buhlmann2014cam}. Point processes, however, are fundamentally different. They are doubly stochastic: the intensity measure governing event generation is itself random, and conditional on the intensity, events occur stochastically. Thus, stating that one neuron influences another requires specifying how the random intensity measure of the child process functionally depends on the observed events of its parent processes. The second line is multivariate point process modeling, which predominantly adopts intensity-based parametrization. In particular, Hawkes processes \citep{hawkes1971spectra} model self-exciting dynamics via conditional intensity functions, with extensions incorporating sparsity, low-rank, and non-stationarity \citep{zhou2013learning, cai2024latent, tang2023multivariate}, whereas Cox processes \citep{cox1980point} treat intensities as random quantities. Despite their effectiveness in numerous applications, these approaches typically rely on parametric, often linear, forms of cross-process relationships and can be sensitive to model misspecification. The third line is functional graphical modeling \citep{qiao2019functional, li2018nonparametric, solea2022copula, lee2022functional, lee2023nonparametric}, which is primarily designed for continuous functional data. Point processes, however, violate the core assumptions underlying classical functional data analysis: they consist of discrete and irregular event collections rather than densely observed smooth curves, and do not naturally reside in separable linear Hilbert spaces. Moreover, they often exhibit overdispersion due to clustering and inhibition effects, and standard tools such as spline smoothing, functional principal component analysis, and covariance operators are not directly applicable without substantial modification, since smoothing point processes into continuous functions may destroy their intrinsic event structure.

In this article, we propose a nonparametric structural equation model for multivariate point process data that differs significantly from both intensity-based and classical functional data modeling approaches. Specifically, we model each point process as a Cox process whose random intensity measure depends nonparametrically on its upstream parent processes, forming a structural equation model over a directed acyclic graph \citep{pearl2009causality}. We represent each point process as a random counting measure, treat the entire process as a single object, and model relationships between random measures directly. We then embed these counting measures through a nested two-layer kernel, where the first layer embeds the measures into a reproducing kernel Hilbert space (RKHS) to enable functional data analysis, and the second layer applies a kernel on these embeddings to create a rich space for nonparametric functional approximation. 

We develop an estimation procedure with two main stages. In the first stage, we recover the topological ordering of the directed graph by employing a new variance-to-mean ratio criterion tailored to Cox processes, avoiding classical conditional independence tests or score-based search. Source nodes, which have no parents, behave like Poisson processes and therefore have variance-to-mean ratio equal to one, whereas nodes influenced by unconditioned parents exhibit overdispersion and have ratios greater than one. Since conditioning on all parent processes reduces a Cox process toward its Poisson baseline, we sequentially identify nodes with minimal variance-to-mean ratios as those whose parents have already been selected, thereby progressively recovering the graph ordering. This procedure is computationally efficient, avoids parametric intensity modeling, and is well suited to neural data, where firing variability reflects network-driven activity. In the second stage, given the recovered ordering, we estimate the parent set by developing a regularized process-on-process regression in the embedded RKHS, where both predictors and responses are point processes. We minimize an operator-level least-squares objective with ridge and group penalties to achieve stability and sparsity, and identify parent nodes by thresholding the corresponding functional regression operator norms. Our approach is computationally tractable with guaranteed global solutions, yields explicit operator estimates that enable sign recovery and facilitate biological interpretation, and naturally incorporates group sparsity for parent selection. We further develop a sample-level estimation algorithm under a given coordinate system, and employ Gram matrix truncation to enable efficient computation for large-scale neural recordings involving hundreds or more neurons. 

We also establish rigorous theoretical guarantees. We first derive concentration bounds for empirical covariance operators of measure embeddings by extending functional data analysis theory to point processes. We then prove consistency of topological ordering recovery under minimum separation conditions on the variance-to-mean ratios. Finally, we establish graph recovery consistency under restricted strong convexity, incoherence, and bounded in-degree conditions. Our asymptotic analysis allows the number of processes to grow polynomially with the sample size under bounded in-degree, providing one of the first scalable guarantees for nonparametric graphical models of multivariate point processes.

Our proposal builds on functional SEM \citep{lee2022functional} and point process regression \citep{lee2025gpam}, but addresses a problem neither method resolves: identifying and estimating directed dependencies when both responses and predictors are point processes. Two developments are central to this advance. First, the Cox process structure provides a source of directional identifiability through overdispersion. Whereas \citet{lee2022functional} identify the ordering of smooth random functions through a score criterion that relies on nonlinear parent-child relations, our variance-to-mean ratio criterion identifies the ordering under both linear and nonlinear relations. Thus, the event-generating mechanism itself supplies the information needed to distinguish directions, without requiring nonlinearity. Establishing this result requires concentration bounds for empirical variance-to-mean ratios and an ordering recovery analysis based on minimum separation of these ratios, rather than operator eigenvalues as in \citet{lee2022functional}. Second, we develop process-on-process regression as the basis for directed network estimation. Whereas \citet{lee2025gpam} embed point process predictors to estimate a real-valued additive response function, we also embed the response process, representing each directed edge by a functional regression operator between embedding spaces. This changes both the estimation target and the analytical requirements, leading to an operator-level least-squares formulation and concentration bounds for predictor-response cross-covariance operators of point process embeddings, extending the analysis for random functions in \citet{lee2022functional}. Moreover, the network structure requires both the ordering and parent sets to be learned with a two-stage procedure, a task absent from single-response regression. Our theoretical analysis consequently extends beyond the single-response estimation and selection guarantees of \citet{lee2025gpam}. Taken together, these developments are far from incremental and require several substantially new and different techniques.  

The rest of the article is organized as follows. Section \ref{sec:model} presents the additive Cox SEM. Section \ref{sec:estimation-population} develops the population-level estimation, and Section \ref{sec:estimation-sample} the sample-level algorithm under a coordinate system. Section \ref{sec:theory} establishes the asymptotic theory. Section \ref{sec:simulations} reports the simulations, and Section \ref{sec:application} revisits the motivating neural spike trains study. The Supplementary Appendix collects additional results and technical proofs.

\vspace{-0.1in}
\section{Cox Process Structural Equation Model}
\label{sec:model}

In this section, we formalize our modeling framework to bridge discrete point process and continuous functional space. We first represent point processes as random counting measures. We next introduce our model where each process has a random intensity that depends additively on the upstream processes in a directed acyclic graph. We then map point processes into a reproducing kernel Hilbert space through kernel embedding.

\vspace{-0.1in}
\subsection{Random counting measure and Cox process SEM}

Let $S \subset \mathbb{R}$ be a compact set with Borel $\sigma$-field $\cB(S)$, and $\mathcal{M}(S)$ be the space of finite measures on $(S, \cB(S))$ equipped with the weak topology so that it is a Polish space. Let $\Omega_M \subseteq \mathcal{M}(S)$ be a measurable subset, and $\cF_M$ be the $\sigma$-field generated by the weak topology on $\Omega_M$. Let $(\Omega, \cF, P)$ be a complete probability space.

\vspace{-0.1in}
\begin{definition}[Random Counting Measure]\label{def:rcm}
A random element $M: \Omega \rightarrow \Omega_M$ with distribution $P_M = P \circ M^{-1}$ is called a random measure on $(S, \cB(S))$. 
\vspace{-0.1in}
\end{definition}

A point process can be equivalently represented as a random counting measure. That is, for a realization of a point process $s = \{s_1, \ldots, s_k\} \subset S$, we can associate it with the counting measure $\mu_{s}(B) = \operatorname{card}(s \cap B)$, for $B \in \cB(S)$, where $\operatorname{card}(\cdot)$ is cardinality. This correspondence is bijective. 

For $i \in [p] = \{1, \ldots, p\}$, let $N_i: \Omega \rightarrow \Omega_{N_i}$ denote the $i$th point process, with distribution $P_{N_i} = P \circ N_i^{-1}$, and $\Omega_{N_i} \subseteq \Omega_M$ is the collection of counting measures corresponding to the $i$th process. Let $L^2(P_{N_i})$ be the space of square-integrable functions of $N_i$ equipped with the inner product $\langle f, g \rangle_{L^2(P_{N_i})} = E[f(N_i)g(N_i)]$. Let $\bfN = (N_1, \ldots, N_p) \in \Omega_{\bfN} = \Omega_{N_1} \times \cdots \times \Omega_{N_p}$ denote the $p$-dimensional point process, with distribution $P_{\bfN}$ and product $\sigma$-field $\cF_{\bfN}$.

We next formalize our Cox process SEM. A point process $N$ is a Cox process with the random intensity measure $X \in \Omega_M$, if, conditional on $X$, $N$ is a Poisson process with the intensity $X$. We write it as $N \sim \operatorname{Cox}(X)$, and $E[N | X] = \operatorname{var}[N | X] = X$. Let $\mathsf{G} = ([p], \mathsf{E})$ denote a directed acyclic graph (DAG), with edges $\mathsf{E} \subseteq [p] \times [p]$. For node $i \in [p]$, let $\operatorname{pa}(i) = \{j: (j,i) \in \mathsf{E}\}$ be its parent set, $\mathrm{de}(i)=\{j \in [p]: \text{there is a directed path from } i \text{ to } j \text{ in } \mathsf{G}\}$ be its descendant set, and $\mathrm{nd}(i)=[p]\setminus(\{i\}\cup \mathrm{de}(i))$ be its non-descendant set. Collect all the parent processes as $\bfN_{\operatorname{pa}(i)} = (N_j)_{j \in \operatorname{pa}(i)} \in \bigtimes_{j \in \operatorname{pa}(i)} \Omega_{N_j}$.

\vspace{-0.1in}
\begin{definition}[Cox process SEM]\label{def:cox_sem}
The multivariate point process $\bfN$ follows a Cox process structural equation model with respect to the DAG $\mathsf{G}$ if, for each $i \in [p]$, $N_i \sim \operatorname{Cox}(\alpha_i + \beta_i(\bfN_{\operatorname{pa}(i)}))$, where $\alpha_i \in \Omega_M$ is the baseline intensity measure, and $\beta_i:\bigtimes_{j \in \operatorname{pa}(i)} \Omega_{N_j} \to \Omega_M$ encodes the directional influences of the parent processes.
\end{definition}

Throughout this article, we focus on the additive model, $\beta_i(\bfN_{\operatorname{pa}(i)}) = \sum_{j \in \operatorname{pa}(i)} \widetilde{\beta}_i^j(N_j)$, with the individual main effect $\widetilde{\beta}_i^j: \Omega_{N_j} \rightarrow \Omega_M$ being Lipschitz continuous in the total-variation norm, in that $\|\widetilde{\beta}_i^j(\mu)-\widetilde{\beta}_i^j(\nu)\|_{TV}\le L\|\mu-\nu\|_{TV}$ for some $L<\infty$. This additive structure alleviates the curse of dimensionality while still remaining highly flexible, as shown in Proposition \ref{prop:denseness} later. It thus provides a reasonable balance between statistical efficiency and model expressiveness. The resulting conditional intensity is of the form,
\begin{align}\label{additive-pp}
E[N_i \mid \bfN_{\operatorname{pa}(i)}] = \alpha_i + \sum_{j \in \operatorname{pa}(i)} \widetilde{\beta}_{i}^j(N_j).
\end{align}

\vspace{-0.1in}
\subsection{Nested RKHS via kernel embedding}
\label{subsec:nested RKHS}

We next construct a two-layer nested RKHS via kernel embedding. The first layer embeds random counting measures into a Hilbert space, enabling the application of functional data analysis techniques. The second layer applies another kernel on those embeddings, allowing the model to capture complex relationships among the underlying intensity functions.

We begin with the first-layer kernel and embedding. Let $\kappa: S \times S \rightarrow \mathbb{R}$ be an integrally strictly positive definite (ISPD) and bounded kernel, such that $|\kappa(s, t)| \leq C$ for all $s, t \in S$ with some constant $C > 0$; we take $C = 1$ without loss of generality. Let $\cH_{\kappa}$ be the induced RKHS of $\kappa$, equipped with the inner product $\langle \cdot, \cdot \rangle_{\cH_{\kappa}}$. 

\begin{definition}[Kernel embedding]\label{def: kernel embedding}
The kernel embedding $T$ of the random counting measure of a point process $N$ is the mapping: $\Omega_N\to\cH_{\kappa}$, $N\mapsto T(N) = \int_S \kappa(\cdot, s) \, dN(s) \in \cH_{\kappa}$.
\end{definition}

The embedding $T$ maps counting measures to functions in $\cH_\kappa$, where the inner product corresponds to integrated kernel evaluation: $\langle T(N_i), T(N_{i'}) \rangle_{\cH_\kappa} = \int_S \int_S \kappa(s, t) \, dN_i(s) \, dN_{i'}(t)$, and the value of the embedding at a point process $s = \{s_1, \ldots, s_k\}$ is $T(s) = \sum_{\ell=1}^k \kappa(\cdot, s_\ell)$. The next result summarizes its properties. 

\vspace{-0.1in}
\begin{proposition}\label{T-property}
The kernel embedding $T$ in Definition \ref{def: kernel embedding} satisfies: 
\vspace{-0.05in}
\begin{enumerate}[(a)]
\item Exchangeability: $T[E(N_i \mid \bfN_{\operatorname{pa}(i)})] = E[T(N_i) \mid \bfN_{\operatorname{pa}(i)}]$, $i=1,\dots,p$.
\item Invertibility: $T: \Omega_N\to\cH_\kappa$ is injective, in that $T(N)=T(N')$ implies $N=N'$ for any $N,N'\in\Omega_N$. Equivalently, there exists a unique left inverse $T^{-1}:T(\Omega_N)\to\Omega_N$ satisfying $T^{-1}\circ T = \mathrm{id}_{\Omega_N}$, where $\mathrm{id}_{\Omega_N}$ is the identity operator. 
\end{enumerate}
\end{proposition}

Applying the kernel embedding $T$ on both sides of \eqref{additive-pp} and by Proposition \ref{T-property}, we have
\begin{align}\label{additive-rkhs}
E[T(N_i) \mid \bfN_{\operatorname{pa}(i)}] = T(\alpha_i) + \sum_{j \in \operatorname{pa}(i)} \beta_i^{j}(N_j) = T(\alpha_i) + \sum_{j \in \operatorname{pa}(i)} \beta_i^{j}\circ T^{-1}\circ T(N_j),
\end{align}
where $\beta_i^j=T\circ\widetilde{\beta}_i^j$. Therefore, model \eqref{additive-rkhs} provides an RKHS additive representation that is equivalent to model \eqref{additive-pp}, while offering more tractable analytical properties. 

We next consider the second-layer kernel and embedding that operates on the first-layer embeddings of the predictor point processes. Let $\widetilde{\kappa}: \cH_\kappa \times \cH_\kappa \rightarrow \mathbb{R}$ be a positive definite and bounded kernel. Then $\kappa_{N_i}(N_i,N_i') = \widetilde{\kappa}(T(N_i), T(N_i'))$ forms the nested, or the second-layer kernel for the embedding $T(N_i) \in \cH_\kappa$. This kernel is ISPD, if $\kappa$ is ISPD and $\widetilde{\kappa}$ is a Gaussian kernel. Let $\cH_{N_i}$ be the induced RKHS of $\kappa_{N_i}$, equipped with the inner product $\langle \cdot, \cdot \rangle_{\cH_{N_i}}$. We call $\cH_{N_i}$ the \emph{nested RKHS} of functions of $N_i$ via kernel embedding.

We recognize that $\cH_{N_i}$ is rich enough to approximate any square-integrable function of $N_i$ arbitrarily well, which in turn justifies the flexibility of the additive model \eqref{additive-rkhs}. 

\vspace{-0.1in}
\begin{proposition} \label{prop:denseness}
For the bounded ISPD kernel $\kappa$ and its induced nested kernel $\kappa_{N_i}$,
\vspace{-0.05in}
\begin{enumerate}[(a)]
\item For any finite measure $M_{N_i} \in \mathcal{M}(\Omega_{N_i})$, the space $\mathcal{H}_{N_i}+\mathbb{R}$ is dense in $L^2(M_{N_i})$, and $\bigoplus_{i=1}^p(\mathcal{H}_{N_i}+\mathbb{R})$ is dense in $\bigoplus_{i=1}^p L^2(M_{N_i})$.

\item Additive functions of $\bfN_{\operatorname{pa}(i)}$ with components in the nested RKHSs are dense in the space of all square-integrable additive functions of $\bfN_{\operatorname{pa}(i)}$.
\end{enumerate}
\end{proposition}

There are several choices for the kernels $\kappa$ and $\widetilde{\kappa}$. For $\kappa$, examples include the radial basis function (RBF) kernel, $\kappa(s,t)=\exp(-\gamma_R |s-t|^2)$, and the Laplacian kernel, $\kappa(s,t)=\exp(-\gamma_L |s-t|)$, for $s, t \in S$. For $\widetilde{\kappa}$, examples include the Gaussian kernel, $\widetilde{\kappa}(f, g) = \exp(-\gamma_G \|f - g\|_{\cH_\kappa}^2)$, for $f, g \in \cH_\kappa$. Together, the hierarchical kernel structure transforms point process regression into functional regression, with the response residing in $\cH_\kappa$ and the predictors modeled through $\cH_{N_i}$. 

\vspace{-0.1in}
\section{Population-level Estimation}
\label{sec:estimation-population}

In this section, we develop the population-level estimation procedure, which consists of two main stages. We first recover the topological ordering of the DAG through a variance-to-mean ratio that exploits the overdispersion property of the Cox process. We then identify the parent set through functional regression operators in the embedded space.

\vspace{-0.1in}
\subsection{Ordering identification}
\label{subsec:order}

Let $\pi=(\pi_1,\ldots,\pi_p)$ denote a permutation of $(1, \ldots, p)$ that indicates the directional order $\pi_1 \to \cdots \to \pi_p$ of the DAG. We aim to identify $\pi$ based on nonparametric conditional moment estimation. A key challenge is that the conditional mean and variance generally cannot be estimated consistently. Standard kernel regression estimators suffer from the curse of dimensionality, in that their convergence rates deteriorate exponentially with the dimension of the conditioning variables. To address this challenge, we consider \emph{additive projections} of conditional moments onto a tractable function space. These projections can be consistently estimated using additive kernel regression. More importantly, under our additive Cox process SEM, they coincide with the true conditional moments whenever the conditioning set includes all parent processes and no descendants. Specifically, for a node $i \in [p]$ and a conditioning set $\cC_i \subseteq [p] \setminus \{i\}$, let $\mathcal{A}_{\cC_i}$ be the $L^2$-closure of the space of centered additive functions, $g(\bfN_{\cC_i}) = c + \sum_{j \in \cC_i} g_j(N_j)$, with $c\in\mathbb{R}$ and component functions $g_j\in L^2(P_{N_j})$ satisfying $E[g_j(N_j)]=0$. Let $P_{\cC_i}[Y] = \argmin_{g \in \mathcal{A}_{\cC_i}} E(Y - g(\bfN_{\cC_i}))^2$ denote the $L^2$-projection of $Y\in L^2(\Omega)$ onto $\mathcal{A}_{\cC_i}$. Noting that $E\big[P_{\cC_i}[N_i(S)]\big] = E[N_i(S)]$, we define the \emph{additive variance-to-mean ratio} of node $i$ as,
\vspace{-0.01in}
\begin{align}\label{eqn:variance-mean-ratio}
R_{\cC_i}[N_i] = \frac{E\big(N_i(S) - P_{\cC_i}[N_i(S)]\big)^2}{E\big[P_{\cC_i}[N_i(S)]\big] }.
\end{align}

\noindent
The centering constraint $E[g_j(N_j)]=0$ ensures identifiability of the additive decomposition. The projection $P_{\cC_i}$ uniquely exists because $\mathcal{A}_{\cC_i}$ is a closed linear subspace of $L^2(P_{\bfN_{\cC_i}})$, and it can be estimated by the usual additive kernel regression \citep{hastie2017generalized}. 

We next introduce a key assumption. 

\vspace{-0.1in}
\begin{assumption} \label{ass:a4coxprime}
For $i \in [p]$, and subset $\cC_i\subseteq \mathrm{nd}(i)\setminus\{i\}$, $\operatorname{pa}(i)\not\subseteq \cC_i$, where $\cC_i$ can be $\varnothing$, 
\begin{align*} 
E\!\left[\operatorname{var}\!\left( \sum_{j \in \operatorname{pa}(i)\setminus \cC_i} \widetilde{\beta}_i^j\!\left(N_j\right)(S) \;\Big|\; \bfN_{\cC_i} \right)\right] \;>\; 0.
\end{align*}
\end{assumption}

\noindent
Assumption \ref{ass:a4coxprime} can be viewed as an analogue of the classical \emph{faithfulness} condition in DAG discovery, which requires that every conditional independence in the observed distribution correspond to a separation relation in the underlying graph \citep{spirtes2000causation}. Similarly, Assumption \ref{ass:a4coxprime} requires that whenever the conditioning set $\cC_i$ omits at least one true parent of node $i$, the effect of the omitted parents cannot be fully represented by an additive function of $\bfN_{\cC_i}$. This rules out the degenerate configurations in which the effects of omitted parents vanish, are perfectly predictable from the conditioned processes, or cancel exactly through interactions. Assumption \ref{ass:a4coxprime} imposes a constraint only on a single scalar quantity, i.e., the residual additive variance of $N_i(S)$, rather than on the full collection of conditional independence relationships among the $p$ processes. Besides, it is required only for subsets of the non-descendants of node $i$, rather than uniformly over all subsets of $[p]\setminus\{i\}$. We next state our main ordering identification result.

\vspace{-0.1in}
\begin{theorem}[Ordering identification]
 \label{prop:vmr}
For a node $i\in[p]$, let $\cC_i$ be a conditioning set of its non-descendants. Suppose $0<E[N_i(S)], E[N_i(S)^2]<\infty$, and Assumption \ref{ass:a4coxprime} holds. Then,
\vspace{-0.05in}
\begin{enumerate}[(a)]
\item If $\cC_i\supseteq \operatorname{pa}(i)$, then the ratio $R_{\cC_i}[N_i]=1$; If $\cC_i\not\supseteq \operatorname{pa}(i)$, then the ratio $R_{\cC_i}[N_i]>1$.
\item At step $m$, let $\cS_{m}=\{\pi_1,\dots,\pi_{m-1}\}$ be the set of previously selected nodes, and $\mathcal{U}_m=[p] \setminus \cS_{m}$ be the rest. Then $\min_{i\in\mathcal{U}_m}R_{\cS_{m}}[N_i]=1$, attained precisely by the nonempty set $\{i \in \mathcal{U}_m: \operatorname{pa}(i)\subseteq \cS_{m}\}$ of nodes whose parents have all been selected. \end{enumerate}
\end{theorem}

\noindent
Theorem \ref{prop:vmr} provides a principled way for recovering a topological ordering $\pi=(\pi_1,\ldots,\pi_p)$ of the DAG through the additive variance-to-mean ratio. Specifically, at step $m$, the node $\pi_m$ whose parents are all contained in the previously selected set $\cS_{m}$ attains a ratio of one, whereas every remaining node that has at least one parent outside this set attains a ratio greater than one. This is different from regression-based ordering methods, which assess whether the residual variance decreases as candidate parents are added. Our criterion instead exploits a distinctive property of the additive Cox process: the variance-to-mean ratio equals exactly one if and only if all intensity randomness has been accounted for. This yields a model-free threshold that can be estimated through additive projections. Theorem \ref{prop:vmr} holds regardless of whether the functions $\widetilde{\beta}_i^j$ are linear or nonlinear, because our identifiability arises from the variance-to-mean structure of point process. This is different from \citet{lee2022functional}, where nonlinearity is essential for identifiability.

\vspace{-0.1in}
\subsection{Parent set identification}
\label{subsec:parent-iden}

Given the topological ordering, SEM estimation reduces to the problem of parent set identification, which we formulate as functional regression in the embedded kernel space.

We first introduce a set of covariance operators. For a process $i \in [p]$, recall $\cH_{N_i}$ denotes the nested RKHS induced by the nested kernel $\kappa_{N_i}$. For any subset $U \subseteq [p]$, let $\cH_U = \bigoplus_{i \in U} \cH_{N_i} = \{\textstyle\sum_{i \in U} f_i: f_i \in \cH_{N_i}\}$ denote the additive RKHS, with the inner product $\langle \sum_{i \in U} f_i, \sum_{i \in U} g_i \rangle_{\cH_U} = \sum_{i \in U} \langle f_i, g_i \rangle_{\cH_{N_i}}$. For a singleton $U=\{i\}$, we write $\cH_i$ for $\cH_{N_i}$. Let $f(\bfN_U) = \sum_{i \in U} f_i(N_i)$. We define the following unique bounded linear operators,
\vspace{-0.01in}
\begin{equation*}
\begin{aligned}
& \Sigma_{UV}: \cH_V \to \cH_U, \text{ such that } \langle f, \Sigma_{UV}\, g \rangle_{\cH_U} = \operatorname{Cov}[f(\bfN_U), g(\bfN_V)], \text{ for } U, V \subseteq [p]; \\
& V_{Ui}: \cH_\kappa \to \cH_U, \text{ such that } \langle f, V_{Ui}\,\phi \rangle_{\cH_U} = \operatorname{Cov}[\langle \phi, T(N_i) \rangle_{\cH_\kappa}, f(\bfN_U)], \text{ for } i \in [p], U \subseteq [p] \setminus \{i\}; \\
& \Gamma_{ii}: \cH_\kappa \to \cH_\kappa, \text{ such that } \langle \phi', \Gamma_{ii}\,\phi \rangle_{\cH_\kappa} = \operatorname{Cov}\bigl[\langle \phi', T(N_i) \rangle_{\cH_\kappa}, \langle \phi, T(N_i) \rangle_{\cH_\kappa}\bigr], \text{ for } i \in [p]. 
\end{aligned}
\end{equation*}

We next introduce the functional regression operator that is central to our parent set identification. For node $i=\pi_m$ at step $m$ with the candidate parent set $\cS_m = \{\pi_1, \ldots, \pi_{m-1}\}$, suppose the usual range condition holds, in that $E[T(N_i)\mid\bfN_{\cS_m}]$ lies in the range of the regression operator: there is a Hilbert-Schmidt operator $B:\cH_\kappa\to\cH_{\cS_m}$ such that $\langle\phi,E[T(N_i)\mid\bfN_{\cS_m}]\rangle_{\cH_\kappa}-(B\phi)(\bfN_{\cS_m})$ is almost surely constant for every $\phi\in\cH_\kappa$. This ensures that the equation $\Sigma_{\cS_m \cS_m}\,B = V_{\cS_m i}$ is solvable. For node $i$ and the candidate set $\operatorname{pa}(i) \subseteq \cS_m$, we define the \emph{functional regression operator} (FRO) as
\begin{align} \label{eqn:FRO}
B_{\cS_m \to i}^0 = \Sigma_{\cS_m \cS_m}^{\dagger}\, V_{\cS_m i}, \text{ where } \Sigma_{\cS_m \cS_m}^{\dagger} \text{ is the Moore-Penrose pseudoinverse.}
\end{align}
The next result shows this functional regression operator is the minimum-norm minimizer of an objective function, and further connects it with the problem of parent set identification. 

\vspace{-0.1in}
\begin{theorem}[Parent identification]
\label{prop:faro}
Suppose the same conditions as Theorem \ref{prop:vmr} hold. Then, 
\begin{enumerate}[(a)]
\item The FRO $B_{\cS_m \to i}^0$ is the minimum-norm minimizer of the operator-level objective, 
\vspace{-0.01in}
\begin{align*}
L_i(B) = -2\langle V_{\cS_m i},\, B \rangle_{\mathrm{HS}} + \langle B,\, \Sigma_{\cS_m \cS_m}\, B \rangle_{\mathrm{HS}} = \sum_{a=1}^\infty\operatorname{var}\bigl[\langle\phi_a,T(N_i)\rangle_{\cH_\kappa}-(B\phi_a)(\bfN_{\cS_m})\bigr],
\end{align*}
up to an additive constant, over $\cB_2(\cH_\kappa,\cH_{\cS_m})$, where $\{\phi_a\}_{a=1}^\infty$ is any orthonormal basis of $\cH_\kappa$, and $\mathcal{B}_2$ is the subspace of Hilbert-Schmidt operators.

\item Writing $B_{\cS_m \to i}^0 = \left[(B_{\cS_m \to i}^0)_{j_1}, \ldots, (B_{\cS_m \to i}^0)_{j_{|\cS_m|}}\right]$, with component $(B_{\cS_m \to i}^0)_j:\cH_\kappa\to\cH_j$, then $\|(B_{\cS_m \to i}^0)_j\|_{\mathrm{HS}}>0$ if and only if $j\in\operatorname{pa}(i)$.
\end{enumerate}
\end{theorem}

\noindent
Theorem \ref{prop:faro} establishes a population-level criterion for parent set identification: the parent nodes of node $i$ correspond exactly to the non-zero components of the FRO. The forward implication follows because each genuine parent effect $\widetilde{\beta}_i^{j}(N_j)$ contributes a separate additive component to $E[T(N_i) \mid \bfN_{\cS_m}]$, and Assumption~\ref{ass:a4coxprime} ensures this component is non-degenerate. The reverse implication follows from the minimum-norm property of the pseudoinverse. For any node $j \notin \operatorname{pa}(i)$, $N_j$ does not contribute to $E[T(N_i) \mid \bfN_{\cS_m}]$, and hence has a zero operator component. Moreover, kernel embedding enables efficient nonparametric estimation without suffering from the curse of dimensionality, as each parent's contribution can be estimated separately.

\vspace{-0.1in}
\section{Coordinate Representation and Sample Estimation}
\label{sec:estimation-sample}

In this section, we translate the population-level results into a finite-sample estimation procedure. We first construct a coordinate representation of the embedded point process through Gram matrices, eigen-decomposition, and truncation, which converts the infinite-dimensional covariance operators into finite matrices. We then use these coordinate representations to recover the topological ordering and identify the parent sets. 

\vspace{-0.1in}
\subsection{Coordinate representation}
\label{subsec:coord}

Suppose we observe $n$ independent realizations $\{ N_i^{(k)} \}_{k=1}^n$ for process $i\in[p]$. Let $\mathcal{N}_i^{(k)}=\{s_{i,1}^{(k)},\dots,s_{i,m_{k,i}}^{(k)}\} \subset S$ denote the $k$th point configuration of the $i$th process, where $m_{k,i}=|\mathcal{N}_i^{(k)}|$ is the number of observed events. For an element $f\in\cH$ with basis $\cB$, we write $\lfloor f\rfloor_{\cB}$, or simply $\lfloor f\rfloor$ when the basis is clear, for its coordinate vector. Similarly, for an operator $B:\cH\to\mathcal{K}$ with bases $\cB$ and $\cB'$, we write $\lfloor B\rfloor_{\cB}^{\cB'}$, or simply $\lfloor B\rfloor$, for its coordinate matrix. 

Our coordinate construction proceeds in four steps. We first form empirical Gram matrices for the predictor and response spaces. We next use their leading eigenspaces to obtain low-rank approximations. We then represent each observed point process by its corresponding Karhunen-Lo\`eve (KL) coordinates. Finally, we express the covariance and cross-covariance operators as matrices in these coordinates. Such a representation provides the bridge from the population-level formulation to the finite-sample estimation algorithm.

First, we form empirical Gram matrices from the observed point process data. Note that the predictor and response sides of the functional regression operator reside in \emph{different} Hilbert spaces. Specifically, a predictor process $N_j$ resides in the nested RKHS $\mathcal{H}_{N_j}$, whereas the embedded response $T(N_i)$ resides in the first-layer RKHS $\mathcal{H}_{\kappa}$. We therefore construct separate Gram matrices for the predictor and response sides.

Evaluating the nested kernel $\kappa_{N_j}$ of Section~\ref{subsec:nested RKHS} from observed configurations requires only inner products of the first-layer embeddings $T(N_j^{(k)})=\sum_{\ell=1}^{m_{k,j}}\kappa(\cdot,s_{j,\ell}^{(k)})\in\cH_\kappa$, i.e., 
\begin{align}\label{eq:embedding_ip}
\bigl\langle T(N_j^{(k)}),\, T(N_j^{(\ell)})\bigr\rangle_{\cH_\kappa} = \sum_{a=1}^{m_{k,j}} \sum_{b=1}^{m_{\ell,j}} \kappa\!\bigl(s_{j,a}^{(k)},\, s_{j,b}^{(\ell)}\bigr).
\end{align}
From \eqref{eq:embedding_ip}, the second-layer nested kernel follows as
\vspace{-0.1in}
\begin{equation}\label{eq:nested_kernel_eval}
\kappa_{N_j}\!\bigl(N_j^{(k)},N_j^{(\ell)}\bigr) = \exp\!\Bigl(-\gamma_G\bigl\|T(N_j^{(k)})-T(N_j^{(\ell)})\bigr\|_{\cH_\kappa}^2\Bigr),
\vspace{-0.1in}
\end{equation}
where the squared distance is evaluated using inner products of the form in \eqref{eq:embedding_ip}. 

For each predictor process $j$, we form the Gram matrix and its centered version as
\vspace{-0.1in}
\begin{equation} \label{eq:predictor_gram}
[K_j^P]_{k,\ell} = \kappa_{N_j}(N_j^{(k)},N_j^{(\ell)}), \qquad G_j^P = Q_n K^P_{j} Q_n
\vspace{-0.1in}
\end{equation}
where $Q_n=I_n-n^{-1}\mathbf{1}_n\mathbf{1}_n^{T}$. Centering replaces each feature map by its deviation $\bar\kappa_{N_j}(\cdot,N_j^{(k)})$ from the empirical kernel mean embedding $\widehat\mu_j=n^{-1}\sum_{k=1}^{n}\kappa_{N_j}(\cdot,N_j^{(k)})$. 

For each response process $i$, we likewise form the first-layer response Gram matrix as
\vspace{-0.1in}
\begin{equation} \label{eq:response_gram}
[K_i^R]_{k,\ell}=\bigl\langle T(N_i^{(k)}), T(N_i^{(\ell)})\bigr\rangle_{\cH_\kappa}, \qquad G_i^R = Q_n K_i^R Q_n.
\vspace{-0.1in}
\end{equation}
These Gram matrices are computed once and reused throughout the estimation procedure.

Next, we use the leading eigenspaces of Gram matrices to obtain low-rank approximations. Specifically, we eigen-decompose the predictor and response Gram matrices,
\vspace{-0.05in}
\begin{equation} \label{eq:eigen_gram}
G_j^P = U^P_{j}\,D_{j}^P (U^P_{j})^{T} + U^{P0}_{j} D^{P0}_{j} (U^{P0}_{j})^{T}, \quad G_i^R=U_i^R D_i^R (U_i^R)^{T}+U_i^{R0} D_i^{R0} (U_i^{R0})^{T}, 
\vspace{-0.05in}
\end{equation}
retaining only the leading $d_j^P$ eigen-pairs for the predictor matrices, $d_i^R$ eigen-pairs for the response matrices, and discarding the rest. This low-rank approximation is particularly useful for reducing the computational cost. 

Then, we represent each observed point process using the corresponding KL coordinates. For predictor process $j$, the coordinate of realization $k$ along the $a$th eigenfunction is 
\vspace{-0.1in}
\begin{equation} \label{eq:kl_coeff}
\widehat{\alpha}_{k,a}^{(j)}=(D^P_{j})_a^{1/2}\,(U^P_{j})_{k,a}, \qquad k\in[n],\; a\in[d^P_j],
\vspace{-0.1in}
\end{equation}
We collect these coordinates into the vector $A_{N_j}^{(k)}=(\widehat{\alpha}_{k,1}^{(j)},\dots,\widehat{\alpha}_{k,d^P_j}^{(j)})^{T}$. For a candidate set $\cS_m$, we form $A_{\cS_m}^{(k)}\in\mathbb{R}^{d_{\cS_m}}$ by stacking $A_{N_{\pi_1}}^{(k)}, \ldots, A_{N_{\pi_{m-1}}}^{(k)}$, where $d_{\cS_m}=\sum_{j\in\cS_m} d^P_j$. Analogously, for response process $i$, the coordinate of realization $k$ along $\widehat{f}_i^a$ is 
\vspace{-0.1in}
\begin{equation} \label{eq:response_kl}
\widehat{\theta}_{k,a}^{(i)} =\bigl\langle T(N_i^{(k)})-\bar{T}_i,\, \widehat{f}_i^a\bigr\rangle_{\cH_\kappa} =(D_i^R)_a^{1/2}\,(U_i^R)_{k,a}, \qquad k\in[n],\; a\in[d_i^R],
\vspace{-0.05in}
\end{equation}
where $\bar{T}_i=n^{-1}\sum_k T(N_i^{(k)})$. We collect these coordinates into $\Theta_{N_i}^{(k)}=(\widehat{\theta}_{k,1}^{(i)},\ldots,\widehat{\theta}_{k,d_i^R}^{(i)})^{T}$.

Finally, we express the covariance operators as matrices in these coordinates. For simplicity, we set $d_j^P = d$ and $d_i^R = d'$. Then, for a candidate set $\cS_m$ and response node $i$, the rank-$d$ truncated sample covariance operators have coordinate matrices
\vspace{-0.01in}
\begin{equation}\label{eq:sample_cov_coord}
\begin{aligned}
& \lfloor\widehat{\Sigma}_{\cS_m \cS_m}^d\rfloor = \frac{1}{n}\sum_{k=1}^n A_{\cS_m}^{(k)}\,(A_{\cS_m}^{(k)})^{T} \in\mathbb{R}^{d_{\cS_m}\times d_{\cS_m}}, \;\; 
\lfloor\widehat{V}_{\cS_m i}^d\rfloor = \frac{1}{n}\sum_{k=1}^n A_{\cS_m}^{(k)}\,(\Theta_{N_i}^{(k)})^{T} \in\mathbb{R}^{d_{\cS_m}\times d'}, \\
&\lfloor\widehat{\Gamma}_{ii}^d\rfloor = \frac{1}{n}\sum_{k=1}^n \Theta_{N_i}^{(k)}\,(\Theta_{N_i}^{(k)})^{T} = n^{-1}D_i^R \in\mathbb{R}^{d'\times d'}.
\end{aligned}
\end{equation}

\vspace{-0.1in}
\subsection{Estimation algorithm}
\label{subsec:algorithm}

With the coordinate representation in hand, we now develop a finite-sample procedure for DAG estimation. Following the population-level results in Section \ref{sec:estimation-population}, the procedure consists of two stages. We first recover a topological ordering by sequentially evaluating an empirical additive variance-to-mean criterion based on Theorem \ref{prop:vmr}. Then, given the estimated ordering, we identify the parents of each node by fitting a block-sparse functional regression model based on Theorem \ref{prop:faro}. Algorithm \ref{alg:main} summarizes the complete procedure. 

First, we recover the topological ordering. At the initial step, we identify a source node, i.e., the node with no parents, using the marginal variance-to-mean ratio. Specifically, for each node $i \in [p]$, we estimate the ratio by
\begin{align}\label{eq:vmr_marginal}
\widehat{R}_{\emptyset}[N_i] = \frac{\sum_{k=1}^n (N_i^{(k)}(S) - \bar{N}_i)^2}{(n-1)\,\bar{N}_i}, \qquad \bar{N}_i = \frac1n\sum_{k=1}^n N_i^{(k)}(S).
\end{align}
At the population level, by Theorem \ref{prop:vmr}, this ratio equals one for a source node and exceeds one for a non-source node. 
 After the first node has been selected, we evaluate each remaining node conditionally on the previously selected nodes. At step $m$, let $\cS_m=\{\widehat{\pi}_1,\dots,\widehat{\pi}_{m-1}\}$ and $\cU_{m} = [p] \setminus \cS_m$ denote the set of previously selected nodes and the set of remaining nodes, respectively. For each $i \in \cU_m$, we estimate the projection in \eqref{eqn:variance-mean-ratio} by additive kernel ridge regression in the nested RKHSs, using $\kappa_{N_j}$ as the reproducing kernel for the component associated with $N_j$, i.e., $\widehat{P}_{\cS_m}[N_i(S)] = \widehat{c} + \sum_{j\in\cS_m}\widehat{g}_j(N_j)$, where $\widehat{c}$ and $\{\widehat{g}_j\}$ solve
\begin{align}\label{eq:additive_regression}
\min_{c\in\mathbb{R},\; g_j\in\cH_{N_j}} \left\{ \frac1n\sum_{k=1}^n \left( N_i^{(k)}(S)-c-\textstyle\sum_{j\in\cS_m}g_j(N_j^{(k)}) \right)^2 + \lambda_{n,1} \sum_{j\in\cS_m}\|g_j\|_{\cH_{N_j}}^2 \right\},
\end{align}
subject to the empirical centering constraint $\sum_{k=1}^{n} g_j(N_j^{(k)})=0$, which we impose by evaluating each $g_j$ through the centered feature map, $g_j(N_j^{(k)})=\langle g_j,\kappa_{N_j}(\cdot,N_j^{(k)})-\widehat\mu_j\rangle_{\cH_{N_j}}$. By the representer theorem, this optimization depends on the observed predictor processes only through their centered Gram matrices. Writing $G^{\mathrm{add}}=\sum_{j\in\cS_m}G^P_{j}$, the fitted values at the sample points are $\bar{N}_i\mathbf{1}_n+G^{\mathrm{add}}(G^{\mathrm{add}}+n\lambda_{n,1} I_n)^{-1}Q_n (N_i^{(1)}(S),\dots,N_i^{(n)}(S))^{T}$. The corresponding empirical additive variance-to-mean ratio is 
\vspace{-0.01in}
\begin{align}\label{eq:vmr_conditional}
\widehat{R}_{\cS_m}[N_i] = \frac{1}{n\,\bar{N}_i}\sum_{k=1}^n \left( N_i^{(k)}(S)-\widehat{P}_{\cS_m}[N_i(S)](\bfN_{\cS_m}^{(k)}) \right)^2.
\end{align}
Here, the marginal ratio in \eqref{eq:vmr_marginal} uses the usual unbiased sample variance, whereas the conditional ratio in \eqref{eq:vmr_conditional} is defined through the empirical mean squared projection residual, matching its population counterpart. 

We then construct the ordering sequentially. Starting with $\cS_1=\emptyset$ and $\mathcal{U}_1=[p]$, at each step $m=1,\ldots,p$, we select the node and update
\vspace{-0.1in}
\begin{equation} \label{eq:updateSU}
\widehat{\pi}_m=\argmin_{i\in\cU_m} \left| \widehat{R}_{\cS_m}[N_i]-1 \right|, \quad \cS_{m+1} = \cS_m \cup \{ \widehat{\pi}_m \}, \quad \cU_{m+1}=\cU_m\setminus\{\widehat{\pi}_m\}.
\end{equation}
By Theorem \ref{prop:vmr}, the correct next node in a valid topological ordering attains a population ratio of one after conditioning on the previously selected nodes. The sequential criterion therefore yields an estimated topological ordering.

Next, given the estimated ordering, we identify the parents of each node. For node $i = \widehat{\pi}_m$, $m\ge2$, all potential parents belong to its predecessor set $\cS_m=\{\widehat{\pi}_1,\dots,\widehat{\pi}_{m-1}\}$. We estimate the functional regression operator $B_{\cS_m\to i}^0$ in \eqref{eqn:FRO} using the coordinate representation of Section \ref{subsec:coord}. Let $\lfloor B_i\rfloor \in \mathbb{R}^{d_{\cS_m}\times d_i^R}$ denote its coordinate matrix, partitioned into predictor-specific blocks $\lfloor(B_i)_j\rfloor\in\mathbb{R}^{d_j^P \times d_i^R}$, $j\in\cS_m$. Substituting the truncated coordinate operators in \eqref{eq:sample_cov_coord} into the operator-level objective of Theorem~\ref{prop:faro} gives the empirical loss,
\begin{align}\label{eq:coord_objective}
\widehat{L}_i(\lfloor B_i\rfloor) = -2\,\mathrm{tr}\bigl(\lfloor\widehat{V}_{\cS_m i}^d\rfloor^{T}\lfloor B_i\rfloor\bigr) + \mathrm{tr}\bigl(\lfloor B_i\rfloor^{T}\lfloor\widehat{\Sigma}_{\cS_m \cS_m}^d\rfloor\lfloor B_i\rfloor\bigr).
\end{align}
To obtain a stable and block-sparse estimator, we minimize the penalized objective
\begin{align}\label{eq:penalized_objective}
\widehat{PL}_i(\lfloor B_i\rfloor) = \widehat{L}_i(\lfloor B_i\rfloor) + \lambda_{n,2} \|\lfloor B_i\rfloor\|_F^2 + \lambda_{n,2} \sum_{j\in\cS_m}\|\lfloor(B_i)_j\rfloor\|_F,
\end{align}
where the first ridge penalty term stabilizes estimation, and the second group penalty term encourages entire predictor-specific blocks to be zero. For tuning simplicity, we use the same penalty parameters $\lambda_{n,2}$. Because each block corresponds to one potential parent process, block sparsity translates directly into edge selection. The objective in \eqref{eq:penalized_objective} is convex and can be solved using disciplined convex programming \citep{boyd2004convex}. The estimated parent set is $\widehat{\mathrm{pa}}(i)=\{j\in\cS_m:\|\lfloor(\widehat{B}_i)_j\rfloor\|_F>0\}$, which is the empirical counterpart of the population representation in Theorem \ref{prop:faro}. 

We note that, when $p$ is large relative to $n$, we may restrict the conditioning set in \eqref{eq:additive_regression} to $\cS_m\cap\widehat{\mathsf{N}}_i$, where $\widehat{\mathsf{N}}_i\subseteq[p]\setminus\{i\}$ is a pre-selected neighborhood containing all parents of node $i$. Such neighborhoods can be obtained by minimizing \eqref{eq:penalized_objective} over $[p]\setminus\{i\}$, in place of~$\cS_m$. In subsequent analysis, we bound the conditioning set sizes by $M=\max_{i\in[p]}|\widehat{\mathsf{N}}_i|$.

Algorithm~\ref{alg:main} summarizes the full estimation procedure. 

\begin{algorithm}[t!]
\caption{DAG learning for multivariate point processes.}
\label{alg:main}
\begin{algorithmic}
\STATE \textbf{Input}: Observed point processes $\{\bfN^{(k)}\}_{k=1}^n$; kernels $\kappa,\widetilde{\kappa}$; regularization parameters $\lambda_{n,1}, \lambda_{n,2}$; truncation levels $\{d^P_j\} = d$, $\{d_i^R\} = d'$.

\STATE \textbf{Coordinate construction}: For each predictor process $j\in[p]$, compute $G_j^P$ via \eqref{eq:predictor_gram}, eigen-decompose it via \eqref{eq:eigen_gram}, and store the coefficients $\{A_{N_j}^{(k)}\}$ via \eqref{eq:kl_coeff}. For every response process $i\in[p]$, compute $G_i^R$ via \eqref{eq:response_gram}, eigen-decompose it via \eqref{eq:eigen_gram}, and store $\{\Theta_{N_i}^{(k)}\}$ via~\eqref{eq:response_kl}.

\STATE \textbf{Ordering recovery}: Initialize $\cS_1=\emptyset$ and $\mathcal{U}_1=[p]$. For $m=1,\ldots,p$ and each $i\in\cU_m$, compute $\widehat{R}_{\cS_m}[N_i]$ via \eqref{eq:vmr_marginal} when $m=1$; or via \eqref{eq:vmr_conditional} when $m>1$. Select $\widehat{\pi}_m$, and update $\cS_{m+1}$ and $\cU_{m+1}$ via \eqref{eq:updateSU}.

\STATE \textbf{Parent identification}: Initialize $\widehat{\mathsf{E}}=\emptyset$. For $m=2,\dots,p$, set $i\leftarrow\widehat{\pi}_m$, $\cS_m\leftarrow\{\widehat{\pi}_1,\dots,\widehat{\pi}_{m-1}\}$. Solve $\lfloor\widehat{B}_i\rfloor$ by minimizing \eqref{eq:penalized_objective}. Add $\left\{ (j,i):\|\lfloor(\widehat{B}_i)_j\rfloor\|_F>0 \right\}$ to $\widehat{\mathsf{E}}$.

\STATE \textbf{Output}: Estimated DAG $\widehat{\mathsf{G}}=([p],\widehat{\mathsf{E}})$.
\end{algorithmic}
\end{algorithm}

The computational cost of our method is dominated by the ordering step, which requires up to $p^2$ additive kernel ridge regressions, each involving an $n \times n$ system. Letting $\bar{m}$ denote the average number of events per realization, the overall complexity is $O(p^2 n^3 + pn^2\bar{m}^2)$. Rank-$d$ truncation further reduces the cost by performing regression in a lower-dimensional coordinate space, with $d \ll n$. Unlike the factorial search over graph orderings or the combinatorial search in methods such as PC \citep{spirtes2000causation} and GES \citep{chickering2002optimal}, our algorithm scales polynomially, making it feasible for moderately large networks.

\vspace{-0.1in}
\section{Asymptotic Theory}
\label{sec:theory}

In this section, we establish the asymptotic guarantees for the proposed estimation procedure. We first introduce the regularity conditions and derive concentration bounds for the empirical operators. We then establish consistency of topological ordering and parent selection, and combine these results to obtain an end-to-end guarantee for DAG recovery.

\vspace{-0.1in}
\subsection{Regularity conditions}
\label{subsec:assumptions}

\begin{assumption}[Minimum signal strength] \label{ass:functions}
For every $i \in [p]$ with $\operatorname{pa}(i) \neq \emptyset$, the true parent components satisfy that $\min_{j \in \mathrm{pa}(i)} \|(B_{\cS_m \to i}^0)_j\|_{\mathrm{HS}} \geq c_{\min}$, for some constant $c_{\min} > 0$.
\end{assumption}

\begin{assumption}[Operator condition] \label{ass:operators}
For any subsets $U, V \subseteq [p]$, define the correlation operator $C_{UV}: \cH_V \to \cH_U$ by $D_U C_{UV} D_V = \Sigma_{UV}$, where $D_U$ is the diagonal operator with $[D_U]_{j,j} = \Sigma_{jj}^{1/2}$ for $j \in U$. For each node $i \in [p]$ with $\operatorname{pa}(i) \neq \emptyset$, write $B_i^0=B^0_{\cS_m\to i}$, let $F_i^0$ be the diagonal operator on $\cH_{\operatorname{pa}(i)}$ with $[F_i^0]_{j,j} = \{1+1/(2\|(B_i^0)_j\|_{\mathrm{HS}})\}\,I$ for $j \in \operatorname{pa}(i)$, and let $H_i^0$ be an operator satisfying $B_i^0 = (F_i^0)^{-1}\,D_{\operatorname{pa}(i)}C_{\operatorname{pa}(i)\operatorname{pa}(i)}^{1/2}\,H_i^0$. Then, (a) for the true parent set $\cS = \mathrm{pa}(i)$ and any non-parent $j \in \mathrm{nd}(i) \setminus \cS$, $\|D_j C_{j\cS} C_{\cS\cS}^{-1/2}\|_{\mathrm{op}} \leq (1 - \alpha)\|H_i^0\|_{\mathrm{HS}}^{-1}/2$, for some $\alpha \in (0,1]$; and (b) there exist constants $\rho_{\min},\rho_{\max},c_H>0$, such that $\rho_{\min} I \leq C_{\cS\cS} \leq \rho_{\max} I$ and $\|H_i^0\|_{\mathrm{HS}}\leq c_H|\cS|^{1/2}$ for every parent set $\cS=\operatorname{pa}(i)\neq\emptyset$.
\end{assumption}

\begin{assumption}[Graph structure] \label{ass:graph}
The DAG satisfies (a) bounded maximum in-degree: $q = \max_{i \in [p]} |\mathrm{pa}(i)| < \infty$; and (b) edge sparsity: the number of edges satisfies $|\mathsf{E}| = O(p)$.
\end{assumption}

\begin{assumption}[Truncation regularity] \label{ass:truncation}
Let $\{\lambda_j^a\}$ and $\{\gamma_i^a\}$ be the eigenvalues of $\Sigma_{jj}$ and $\Gamma_{ii}$ in nonincreasing order, and $\{\eta_j^a\}$ the corresponding eigenfunctions of $\Sigma_{jj}$. Then, (a) Eigenvalue gaps: $\tau_d:=\min_{j\in[p]}(\lambda_j^d-\lambda_j^{d+1})>0$, and $\min_{i\in[p]}(\gamma_i^{d'}-\gamma_i^{d'+1})\ge\tau_d$; (b) Truncation approximation: Let $P_j^d$ be the projection onto $\cH^d_{N_j}=\mathrm{span}\{\eta^a_{j}: a\in[d]\}$, and $Q_i^{d'}$ the projection onto the span of the leading $d'$ eigenfunctions of $\Gamma_{ii}$. Then, $\max_{i\in[p]} \; \max_{j\in\operatorname{pa}(i)} \bigl\|P_j^d\,(B^0_i)_j Q_i^{d'} - (B^0_i)_j\bigr\|_{\mathrm{HS}} \;\le\; C_{\mathrm{trunc}}\, d^{-\nu}$, for some $C_{\mathrm{trunc}}\ge0$, $\nu>0$, where $(B^0_i)_j:\cH_\kappa\to\cH_j$ is the $j$th component of the FRO $B^0_i:=B^0_{\cS_m\to i}$.
\end{assumption}

\noindent
Assumption~\ref{ass:functions} is a minimum-signal condition requiring each true parent to contribute a detectable effect. Such a condition is essential for finite-sample parent selection and is the functional analogue of the beta-min conditions standard in high-dimensional variable selection. Assumption \ref{ass:operators}(a) imposes group incoherence for lasso-type selection consistency, paralleling Assumption 8 of \citet{lee2022functional}, whereas \ref{ass:operators}(b) bounds the spectrum of the correlation operator of each parent set away from zero and infinity and controls the norm of $H_i^0$, ensuring stable inversion and controlling the regularization bias in parent selection. Both parts use correlation rather than covariance operators because trace-class covariance eigenvalues accumulate at zero, precluding a uniform positive lower bound on $\Sigma_{\cS\cS}$ in infinite dimensions. Normalization removes marginal scale effects and permits meaningful operator-level conditions. The weight $F_i^0$ in Assumption \ref{ass:operators}(a) reflects the ridge and group penalties in \eqref{eq:penalized_objective}, and the factorization through $C_{\cS\cS}^{1/2}$ keeps it valid at any truncation level. Assumption \ref{ass:graph} imposes bounded in-degree and edge sparsity, allowing the sample complexity of parent selection to depend polynomially on the maximum in-degree $q$ but only polylogarithmically on the ambient dimension $p$, consistent with sparse connectivity commonly observed in neural and other biological networks. Assumption~\ref{ass:truncation} governs the finite-dimensional Karhunen-Lo\`eve approximation used in estimation. It imposes a positive eigengap to ensure Davis-Kahan stability of the estimated eigen-spaces, and an operator-approximation rate of $d^{-\nu}$. These conditions are all standard in functional principal component analysis.

\vspace{-0.1in}
\subsection{Operator concentration bounds}
\label{subsec:bounds}

We first derive concentration bounds for the empirical operators. The empirical counterparts of the population-level operators $\Sigma_{jk}$, $V_{ji}$, and $\Gamma_{ii}$ are given by 
\begin{equation*}
\begin{aligned}
& \widehat\Sigma_{jk}=\frac1n\sum_{\ell=1}^n\bar\kappa_{N_j}(\cdot,N_j^{(\ell)})\otimes\bar\kappa_{N_k}(\cdot,N_k^{(\ell)}), \qquad\quad
\widehat V_{ji}=\frac1n\sum_{\ell=1}^n\bar\kappa_{N_j}(\cdot,N_j^{(\ell)})\otimes[T(N_i^{(\ell)})-\bar T_i], \\
& \widehat\Gamma_{ii}=\frac1n\sum_{\ell=1}^n[T(N_i^{(\ell)})-\bar T_i]\otimes[T(N_i^{(\ell)})-\bar T_i], \quad \text{ where } \;\; \bar{T}_i = n^{-1}\sum_{\ell=1}^n T(N_i^{(\ell)}).
\end{aligned}
\end{equation*}
The population-level rank-$d$ truncated operators $\Sigma^d_{jk}, V^d_{ji}, \Gamma^d_{ii}$ are given by 
\begin{equation*}
\begin{aligned}
& \Sigma^d_{jk} = \sum_{a,b\in[d]} E[\alpha_a^{(j)}\alpha_b^{(k)}]\,\eta_j^a\otimes\eta_k^b, \qquad 
V^d_{ji} = \sum_{a\in[d],\,b\in[d']} E[\alpha_a^{(j)} \theta_b^{(i)}]\,\eta_j^a\otimes f_i^b, \\
& \Gamma^d_{ii} = \sum_{a,b\in[d']} E[\theta_a^{(i)}\theta_b^{(i)}]\,f_i^a\otimes f_i^b,
\end{aligned}
\end{equation*}
where $\{\eta_j^a\}$ and $\{f_i^b\}$ are eigenfunctions, with corresponding KL coefficients $\alpha_a^{(j)}$ and $\theta_b^{(i)}$, of $\Sigma_{jj}$ and $\Gamma_{ii}$, respectively. The sample-level truncated operators $\widehat{\Sigma}^d_{jk},\widehat{V}^d_{ji},\widehat{\Gamma}^d_{ii}$ are given in \eqref{eq:sample_cov_coord}. Because the truncation levels are small relative to $n$, this coordinate representation replaces operations on full $n\times n$ Gram matrices with low-dimensional matrix operations. The next theorem establishes uniform concentration bounds for the basic empirical operators.

\vspace{-0.1in}
\begin{theorem}[Concentration of covariance operators]
\label{thm:concentration}
Let $\mu_j=E[\kappa_{N_j}(\cdot,N_j)]$. There exist positive constants $C_1,\ldots,$ $C_8$, such that, for any $u>0$,
\vspace{-0.1in}
\begin{enumerate}[(a)]
\setlength{\itemsep}{1pt}
\setlength{\parsep}{0pt}
\item Mean element: $\mathbb{P}\left( \max_{j\in[p]}\|\widehat\mu_{j}-\mu_{j}\|_{\cH_{N_j}}\ge u \right) \le 2\exp\{-C_1 n(u\wedge u^2)+\log p\}$. 

\item Response covariance: $\mathbb{P}\left( \max_{i\in[p]}\|\widehat\Gamma_{i i}-\Gamma_{i i}\|_{\mathrm{HS}}\ge u \right) \le C_3\exp\{-C_4 (nu^2\wedge (nu)^{1/2}) + \log p\}$. 

\item Truncated predictor covariance: $\mathbb{P}\left( \max_{j,k\in[p]}\|\widehat\Sigma^d_{j k}-\Sigma^d_{j k}\|_{\mathrm{HS}}\ge u \right) \le C_5\exp\{-C_6 \, n(u\wedge u^2) \tau_d^2 + \log p\}$.

\item Truncated predictor-response cross-covariance: $\mathbb{P}\left( \max_{j,i\in[p]}\|\widehat{V}^d_{j i} - V^d_{j i}\|_{\mathrm{HS}} \ge u \right) \le C_7\exp\{-C_8 \, (nu^2\wedge (nu)^{1/2}) \tau_d^2 + \log p\}$. 
\end{enumerate}
\end{theorem}

\noindent 
The bounds in Theorem \ref{thm:concentration} hold uniformly over all nodes, with the $\log p$ term reflecting the cost of uniform control. Theorem \ref{thm:concentration}(a) and (b) follow from Bernstein-type inequalities under the bounded-kernel condition, while Theorem \ref{thm:concentration} (c) and (d) account for estimation of the truncated eigenspaces. Consequently, their rates depend on the truncation level $d$ and the minimum eigengap $\tau_d$ through the Davis-Kahan perturbation bound. 

\vspace{-0.1in}
\subsection{Consistency of ordering, parent selection, and DAG recovery}
\label{subsec:main_results}

We next establish consistency of ordering identification, parent selection, and DAG recovery, while allowing $p$ and $d$ to grow with $n$.

\vspace{-0.1in}
\begin{theorem}[Ordering identification consistency]
\label{thm:ordering}
Suppose Assumption \ref{ass:a4coxprime} holds. Suppose, for every $i\in[p]$ and $\cC\subseteq\mathrm{nd}(i)$, the additive projection in~\eqref{eqn:variance-mean-ratio} is $P_{\cC}[N_i(S)]=E[N_i(S)]+g(\bfN_{\cC})-E[g(\bfN_{\cC})]$ for some $g\in\cH_{\cC}$ with $\|g\|_{\cH_{\cC}}\lesssim1$ uniformly in $i$ and $\cC$. Let $\widehat\pi$ denote the estimated topological ordering from Algorithm~\ref{alg:main}. Define the minimum separation $\Delta_{\min} = \min_{i\in[p],\,\cC\subseteq\mathrm{nd}(i):\,\operatorname{pa}(i)\not\subseteq\cC} \bigl\{R_{\cC}[N_i]-1\bigr\}>0,$ and the uniform estimation error $\eta_1(M) = \frac{M\sqrt{\log p}}{n^{1/2}} + \lambda_{n,1}$, where $M$ is the bound on the sizes of the conditioning sets in~\eqref{eq:additive_regression}. If $\lambda_{n,1}\gtrsim M\sqrt{\log(p^2/\delta)/n}$, $\eta_1(M)\prec\Delta_{\min}$, and the sample size $n \gtrsim \log^2(p^2/\delta)/\Delta_{\min}$, then, with probability at least $1-\delta$, $\widehat\pi$ is a valid topological ordering of~$\mathsf{G}$, in that $\operatorname{pa}(\widehat\pi_m)\subseteq\{\widehat\pi_1,\ldots,\widehat\pi_{m-1}\}$ for every $m\in[p]$.
\end{theorem}

\noindent
Theorem \ref{thm:ordering} establishes consistency of order identification. The estimation error $\eta_1(M)$ comes from the additive regressions~\eqref{eq:additive_regression} and is uniform over conditioning sets of size at most~$M$. The separation parameter $\Delta_{\min}$ quantifies the minimum population gap by which the variance-to-mean criterion distinguishes a non-source node from a source node. Under Assumption~\ref{ass:a4coxprime}, this gap is strictly positive because any unaccounted-for parent induces extra Poisson variation. The condition $\eta_1(M)\prec\Delta_{\min}$ ensures that the combined sampling and regularization errors are asymptotically smaller than this population gap. Consequently, the empirical criterion preserves the strict inequalities in Theorem~\ref{prop:vmr} at every step of the ordering procedure. This result extends the overdispersion-based ordering principle for Poisson DAGs \citep{park2015learning,park2019high}, as well as the ordering guarantees of \citet{buhlmann2014cam,lee2022functional}, to our point process setting.

\vspace{-0.1in}
\begin{theorem}[Parent selection consistency]
\label{thm:faro_rate}
Suppose Assumptions~\ref{ass:functions}, \ref{ass:operators}, and~\ref{ass:truncation} hold, and let $\delta\in(0,1)$. For a node $i \in [p]$, let $\widehat{\mathrm{pa}}(i)$ be the parent set produced by Algorithm~\ref{alg:main} with a valid topological ordering in place of $\widehat\pi$ and with regularization parameter $\lambda_{n,2} = C_\lambda \tau_d^{-1} \sqrt{\log (p/\delta) /n}$ for a sufficiently large constant $C_\lambda$. If $C_{\mathrm{trunc}}d^{-\nu}\prec\lambda_{n,2}$, and $n \gtrsim (q+\tau_d^{-1})^2\log(p/\delta)/(c_{\min}^2\tau_d^2)+\log^2(p/\delta)/\tau_d$, then,
\begin{enumerate}[(a)]
\setlength{\itemsep}{1pt}
\setlength{\parsep}{0pt}
\item For a constant $C_B>0$, $\mathbb{P}\left(\|\widehat B_i-B^0_i\|_{\mathrm{HS}}>C_B\left\{\frac{|\operatorname{pa}(i)|^{1/2}\lambda_{n,2}}{(\rho_{\min}\tau_d)^{1/2}}+\frac{\|B^0_i\|_{\mathrm{HS}}\log^2(p/\delta)}{n\tau_d}\right\}\right)\le\delta$.

\item $\mathbb{P}\left( \widehat{\mathrm{pa}}(i) = \mathrm{pa}(i) \text{ for all } i \in [p] \right) \geq 1 - \delta$.

\end{enumerate}
\end{theorem}

\noindent
Theorem \ref{thm:faro_rate}(a) establishes convergence of the penalized FRO estimator $\widehat{B}_i$ to $B^0_i$, where the convergence rate, balanced by $\lambda_{n,2} \asymp \tau_d^{-1}(\log p/n)^{1/2}$, depends polynomially on the in-degree \(q\) but only polylogarithmically on the ambient dimension $p$. Its first term reflects estimation on the leading eigenspaces of the parent processes, on which the covariance eigenvalues are bounded below by $\rho_{\min}\tau_d$, and its second term reflects estimation of the response eigenspaces. Theorem \ref{thm:faro_rate}(b) further establishes convergence of parent selection. Despite the infinite-dimensional observations, the required sample size $n \gtrsim q^2\log p \vee\log^2p$ for fixed $d$ is only polylogarithmic in $p$. The factor $q^2$ arises from restricted strong convexity, $c_{\min}^{-2}$ reflects the difficulty of detecting weak effects under Assumption~\ref{ass:functions}, and $\lambda_{n,2}\asymp\tau_d^{-1}\sqrt{\log p/n}$ corresponds to the standard group lasso regularization rate \citep{yuan2006model}. 

\begin{theorem}[Graph recovery consistency]
\label{thm:parents}\label{thm:graph}
Suppose the conditions of Theorems \ref{thm:ordering} and \ref{thm:faro_rate} and Assumption~\ref{ass:graph} hold. Let $\widehat{\mathsf{G}}=([p],\widehat{\mathsf{E}})$ be the estimated DAG.
If 
\begin{align*}
n \geq C\cdot\max\!\left\{ \frac{\log^2(p^2/\delta)}{\Delta_{\min}},\; \frac{(q+\tau_d^{-1})^2\log(p/\delta)}{c_{\min}^2\tau_d^2},\;\frac{\log^2(p/\delta)}{\tau_d} \right\},
\end{align*}
for any $\delta>0$, then, $\mathbb{P}(\widehat{\mathsf{G}}=\mathsf{G})\ge 1-\delta$.
\end{theorem}

\noindent
Theorem~\ref{thm:graph} establishes an end-to-end guarantee for graph recovery, by combining the ordering and parent selection results through a union bound. Conditional on recovering a valid ordering, each node is regressed only on its predecessors and never on its descendants. The resulting sample complexity is polylogarithmic in $p$ and polynomial in $q$, allowing the network size to grow nearly exponentially with $n$ when $M$ is bounded. This result considerably strengthens the functional SEM graph recovery guarantee of \citet{lee2022functional}. 

\vspace{-0.1in}
\section{Simulation Studies}
\label{sec:simulations}

In this section, we evaluate the finite-sample performance of our proposed point process structural equation model (PPSEM) and compare with the count-based alternative.

\vspace{-0.1in}
\subsection{Simulation setup}
\label{subsec:setup}

We generate $n$ independent realizations of a $p$-dimensional Cox process over $S=[0,1]$. We consider four configurations of sample size and network dimension, $(n,p) \in \{(100,50), (100,$ $100), (200,100), (200,200)\}$, with the ratio $n / p \in \{1,2\}$. For each configuration, we generate 20 independent data replications. We consider two graph structures, a hub graph and a random graph. For the hub structure, we partition the $p$ nodes into blocks of ten and connect the first node in each block to the other nine, yielding $0.9p$ edges. For the random structure, we fix an initial node ordering and include each forward edge independently with probability $3/(p-1)$, yielding $3p/2$ edges on average. For both structures, we randomly permute the node labels and generate a new graph in every replication. 

Given a DAG, we generate the point processes in topological order under two models, a linear model and a nonlinear one. Both use baseline intensity $\alpha_i(s)=4.5$ for source nodes and $\alpha_i(s)=1.5$ otherwise. The contribution of parent process $N_j$ to node $i$ for the linear and nonlinear model is, respectively, 
\begin{equation*}
\widetilde\beta_i^j(N_j)(s)
=2\sum_{t\in N_j}\exp\{-12(s-t)^2\}, \qquad \widetilde\beta_i^j(N_j)(s)=\left[\sum_{t\in N_j}\exp\{-12(s-t)^2\}\right]^{1.5},
\end{equation*}
where the superlinear power in the latter amplifies the effect of temporally clustered events and introduces burst-sensitive dependence. We bound the resulting intensity below by $0.3$ in the linear model, and truncate it to $[0.3,20]$ in the nonlinear model. These models assess recovery under both linear effects and nonlinear effects driven by temporal event patterns.

To our knowledge, existing DAG learning methods do not directly accommodate point-process-valued nodes while retaining their event-time information. We therefore compare PPSEM with the Poisson DAG method using moments ratio scoring \citep{park2019high}. Both methods estimate an ordering through a moment-ratio criterion and then select parents by sparse regression. However, Poisson DAG operates on total counts $N_i(S)$, whereas PPSEM uses kernel embeddings of the full counting measures. This provides a useful benchmark for assessing the benefit of retaining event-time information. For PPSEM, we condition on all previously selected nodes, so $M=p-1$. We use a Gaussian kernel with bandwidth $\gamma_R=6$ for the first-layer kernel, and another Gaussian kernel with bandwidth set by the median pairwise embedding distances for the second-layer kernel. We use a common truncation level $d=d'=6$ for all predictor and response coordinates, capturing on average $89.5\%$ of the total variation in the centered Gram matrices.  We select the ordering using the leave-one-out variance-to-mean criterion. We select the regularization parameter $\lambda_{n,1}$ in \eqref{eq:additive_regression} by generalized cross-validation, and select the regularization parameter $\lambda_{n,2}$ in \eqref{eq:penalized_objective} by a BIC-type criterion. For Poisson DAG, we select ordering using moments ratio scoring applied to total counts, followed by lasso Poisson regression for parent selection.

We evaluate DAG recovery using structural Hamming distance (SHD), true discovery rate (TDR), and area under the ROC curve (AUC). SHD counts the edge insertions, deletions, and reversals needed to recover the true DAG; TDR is the proportion of selected edges that are true edges. Both SHD and TDR evaluate the estimated graph under a selected tuning parameter, while AUC summarizes the ROC curve obtained by varying the sparsity parameter. Because each method selects edges only in the forward direction of its estimated ordering, the false positive rate is calculated over eligible non-edges under that ordering, while the true positive rate is calculated over all true edges. Thus, an ordering that excludes true edges limits the attainable true positive rate, even as the sparsity penalty decreases. Lower SHD and higher TDR and AUC indicate better performance. 

\vspace{-0.1in}
\subsection{Simulation results}
\label{subsec:results}

\begin{table}[t!]
\centering
\caption{DAG recovery under linear and nonlinear models, hub and random graphs, and four $(n,p)$ settings. Reported are means (standard deviations) over 20 replications. SHD and TDR evaluate BIC-selected graphs, and AUC summarizes the ROC curve along the sparsity path. Lower SHD and higher TDR and AUC indicate better recovery.}
\label{tab:sims}
\footnotesize
\setlength{\tabcolsep}{3pt}
\begin{tabular}{@{}llccc|ccc@{}}
\toprule
\multirow{2}{*}{Structure} & \multirow{2}{*}{$(n,p)$} & \multicolumn{3}{c}{PPSEM (Ours)} & \multicolumn{3}{c}{Poisson DAG} \\
\cmidrule(lr){3-5}\cmidrule(lr){6-8}
 & & AUC & SHD & TDR & AUC & SHD & TDR \\
\midrule
\multicolumn{8}{c}{Linear model} \\
Hub & (100,50) & 0.91 (0.09) & 56 (36) & 0.55 (0.25) & 0.51 (0.17) & 88 (32) & 0.29 (0.15) \\
 & (100,100) & 0.96 (0.05) & 115 (68) & 0.51 (0.20) & 0.44 (0.16) & 270 (73) & 0.17 (0.09) \\
 & (200,100) & 0.99 (0.01) & 78 (72) & 0.65 (0.26) & 0.51 (0.12) & 226 (68) & 0.22 (0.10) \\
 & (200,200) & 0.99 (0.02) & 126 (117) & 0.69 (0.23) & 0.41 (0.07) & 862 (128) & 0.09 (0.03) \\
\midrule
Random & (100,50) & 0.78 (0.05) & 110 (44) & 0.41 (0.12) & 0.43 (0.08) & 130 (25) & 0.26 (0.07) \\
 & (100,100) & 0.79 (0.04) & 312 (102) & 0.30 (0.07) & 0.38 (0.04) & 326 (58) & 0.19 (0.05) \\
 & (200,100) & 0.83 (0.03) & 173 (62) & 0.48 (0.10) & 0.42 (0.05) & 317 (55) & 0.21 (0.04) \\
 & (200,200) & 0.82 (0.02) & 429 (154) & 0.40 (0.10) & 0.35 (0.03) & 829 (132) & 0.13 (0.03) \\
\midrule
\multicolumn{8}{c}{Nonlinear model} \\
Hub & (100,50) & 0.97 (0.01) & 86 (41) & 0.39 (0.15) & 0.61 (0.16) & 73 (29) & 0.36 (0.15) \\
 & (100,100) & 0.97 (0.04) & 195 (64) & 0.34 (0.12) & 0.52 (0.12) & 203 (59) & 0.24 (0.09) \\
 & (200,100) & 0.99 (0.01) & 141 (94) & 0.48 (0.23) & 0.66 (0.12) & 138 (50) & 0.38 (0.14) \\
 & (200,200) & 0.99 (0.00) & 257 (176) & 0.49 (0.21) & 0.48 (0.07) & 665 (219) & 0.15 (0.07) \\
\midrule
Random & (100,50) & 0.70 (0.06) & 77 (22) & 0.53 (0.12) & 0.41 (0.07) & 121 (26) & 0.28 (0.07) \\
 & (100,100) & 0.69 (0.05) & 160 (53) & 0.52 (0.12) & 0.35 (0.04) & 342 (66) & 0.16 (0.04) \\
 & (200,100) & 0.75 (0.02) & 140 (24) & 0.55 (0.10) & 0.41 (0.04) & 304 (52) & 0.21 (0.05) \\
 & (200,200) & 0.73 (0.03) & 288 (57) & 0.53 (0.10) & 0.34 (0.03) & 857 (156) & 0.12 (0.03) \\
\bottomrule
\end{tabular}
\end{table}

\begin{figure}[t!]
\centering
\begin{tabular}{cc}
(a) Linear model, hub structure & (b) Linear model, random structure \\
\includegraphics[width=0.48\textwidth]{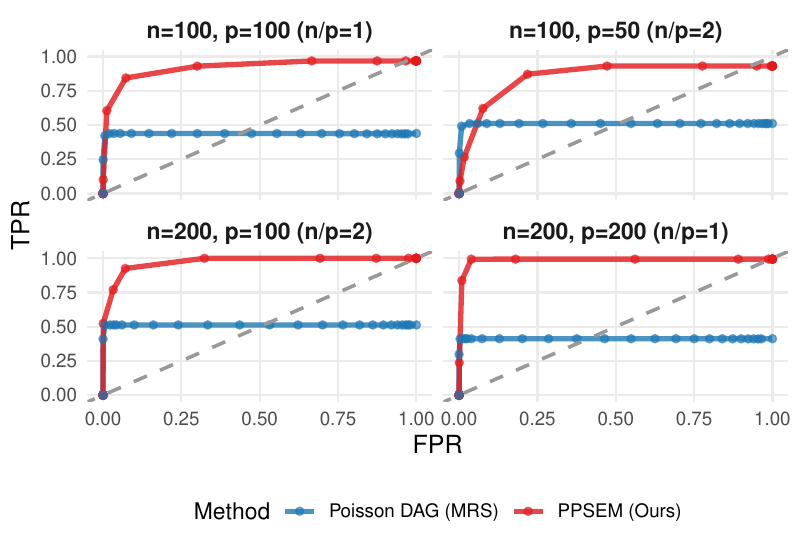} & 
\includegraphics[width=0.48\textwidth]{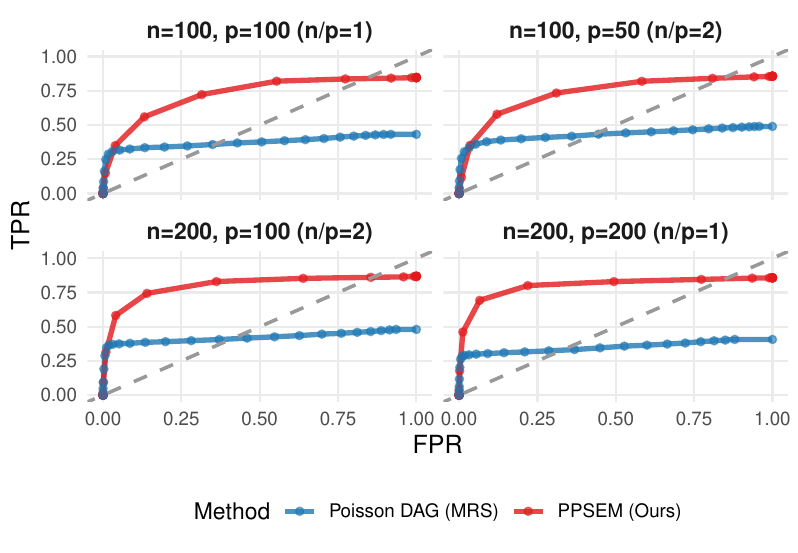} \\
(c) Nonlinear model, hub structure & (d) Nonlinear model, random structure \\
\includegraphics[width=0.48\textwidth]{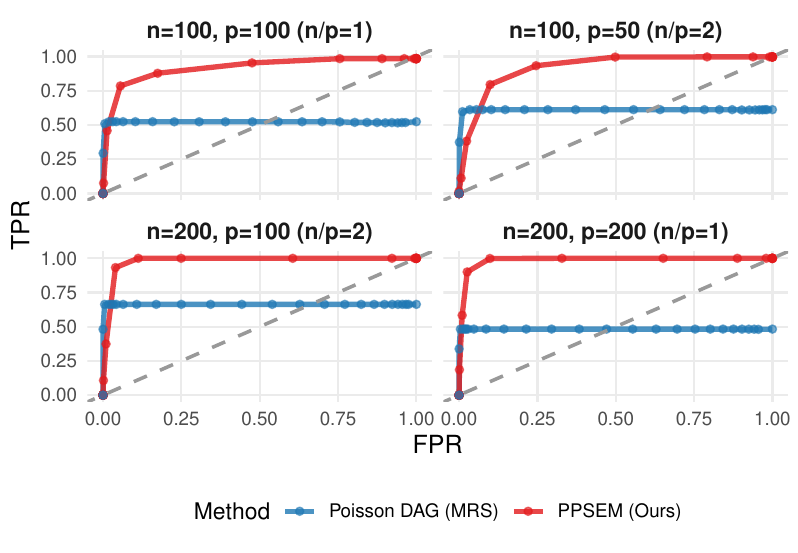} & 
\includegraphics[width=0.48\textwidth]{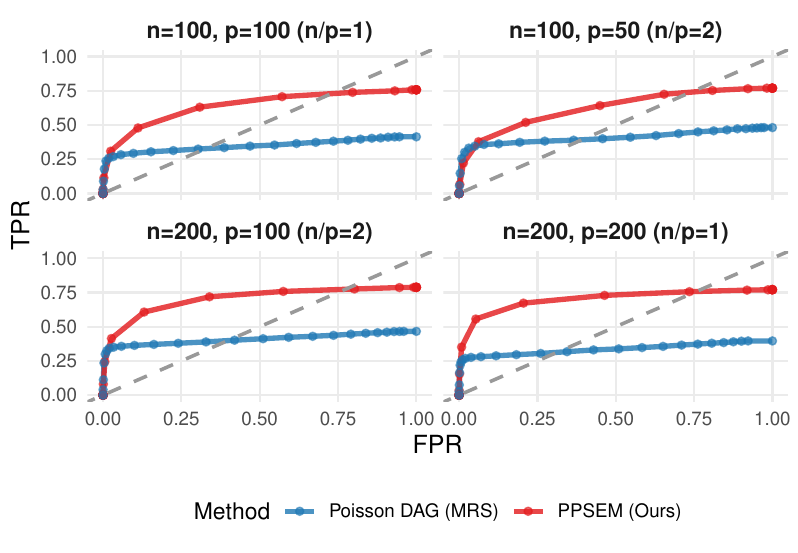} \\
\end{tabular}
\vspace{-0.2in}
\caption{ROC curves for linear and nonlinear models, hub and random graphs, and four $(n,p)$ settings, averaged over 20 replications. Curves trace the sparsity path.} \label{fig:roc_curves}
\end{figure}

Table \ref{tab:sims} reports the results, combining recovery of the BIC-selected graph (SHD and TDR) with performance along the sparsity path (AUC). It is seen that PPSEM achieves higher AUC and TDR in every setting and lower SHD in 14 of the 16 settings. Its advantage is particularly clear for hub graphs: AUC ranges from $0.91$ to $0.99$, compared with $0.41$ to $0.66$ for Poisson DAG. Under the linear model at $(n,p)=(200,200)$, PPSEM reduces SHD from $862$ to $126$ and increases TDR from $0.09$ to $0.69$. These results show that the advantage extends to the graph chosen by the selection criterion, rather than appearing only along the regularization path. Random graphs are more challenging, with PPSEM AUC ranging from $0.69$ to $0.83$, but its advantage over Poisson DAG remains across both generating models and all sample sizes. The two exceptions in SHD occur for nonlinear hub graphs at $(n,p)=(100,50)$ and $(200,100)$, where PPSEM has slightly larger SHD despite higher TDR and AUC. Thus, improved performance along the sparsity path does not invariably translate into lower SHD at the BIC-selected graph. Overall, the table supports an advantage under both linear and nonlinear effects, while identifying the settings in which graph selection remains difficult.

Figure \ref{fig:roc_curves} complements these numerical summaries by showing the recovery trade-off as sparsity varies. PPSEM generally achieves higher true positive rates at comparable false positive rates, indicating that its advantage is not confined to a single selected penalty level. The curves also reach substantially higher true positive levels, consistent with fewer true edges being excluded by the estimated ordering. A low plateau, in contrast, cannot be overcome merely by weakening the parent selection penalty when the corresponding true edges point backward in that ordering. Because the plotted curves cover a finite penalty grid, their endpoints also reflect the explored range of sparsity. 

Together, these results demonstrate the effectiveness of PPSEM, showing that retaining event-time information improves both selected graph recovery and performance across the sparsity path, including under nonlinear, burst-sensitive dependence.

\vspace{-0.1in}
\section{Neural Spike Trains Analysis}
\label{sec:application}

In this section, we apply PPSEM to multiregion neural spike trains recorded during a visual discrimination task \citep{steinmetz2019distributed}, examining how directed dependencies vary with stimulus side and temporal alignment. The same dataset was analyzed by \citet{pinkney2025identifying} using a Bernoulli autoregressive partially linear additive model. PPSEM instead models the spike trains directly as point processes, retaining their event-time information.

\vspace{-0.1in}
\subsection{Data and analysis setup}
\label{subsec:realdata}

We analyze one recording session from a single mouse, available through the DANDI archive (\texttt{sub-Hench\_ses-20170617T120000.nwb}). After excluding trials with missing event times, failed inclusion flags, or equal visual contrast on the two sides, we retain 165 left-stimulus and 119 right-stimulus trials, classified by the side with higher contrast. For each stimulus side, we analyze a $0.6$-second window following either stimulus onset or the recorded wheel-turn response, yielding four conditions. The two alignments use the same underlying trials; within each condition, trials are treated as independent process realizations.

To balance anatomical representation, we select up to four neurons with the highest average firing rates from each recorded region. Eligibility requires at least one spike per trial on average within the analysis window. Regions with fewer than four eligible neurons contribute all available units. Selection is performed once using pooled windows from all four conditions, yielding a common set of $p=32$ neurons. This approach limits the dominance of fast-firing collicular and midbrain neurons while retaining slower-firing cortical neurons. The recorded locations are grouped into visual cortex (VISp, VISl), superior colliculus (SCm, SCsg), midbrain (MB, MRN), and hippocampal areas (CA fields and postsubiculum). One neuron with an unassigned location is retained as a network node but excluded from region-level summaries. It forms a separate group when calculating between-group edge percentages. Further preprocessing details are provided in Appendix Section S2. 

We use a Gaussian kernel for the first-layer and a linear kernel for the second-layer to facilitate interpretation of edge signs. We label associations as excitatory-like or inhibitory-like according to whether increased firing along the predictor neuron's mean profile increases or decreases the response neuron's integrated smoothed activity. To assess sampling stability, we refit the full procedure on 50 subsamples, each containing $80\%$ of the trials drawn without replacement, and retain full-data edges selected in at least half of the refits. Kernel settings, tuning, and the sign calculation are detailed in Appendix Section S2.

\vspace{-0.1in}
\subsection{Estimated neural networks}
\label{subsec:realresults}

\begin{table}[t!]
\centering
\caption{Network summaries for left- and right-stimulus trials aligned to stimulus onset or wheel response. Positive and negative edges denote excitatory-like and inhibitory-like associations. Between-group percentages treat the neuron with an unassigned location as a separate group.}
\label{tab:summary}
\resizebox{\textwidth}{!}{%
\begin{tabular}{lccccccc}
 \toprule
 Dataset & Trial & Dependencies & Excitatory & Inhibitory & Exc (\%) & Inh (\%) & Between (\%) \\
 \midrule
 Stimulus Onset & Left & 36 & 23 & 13 & 63.9 & 36.1 & 75.0 \\
 Stimulus Onset & Right & 63 & 34 & 29 & 54.0 & 46.0 & 68.3 \\
 Wheel Movement & Left & 59 & 30 & 29 & 50.8 & 49.2 & 71.2 \\
 Wheel Movement & Right & 58 & 30 & 28 & 51.7 & 48.3 & 72.4 \\
 \bottomrule
\end{tabular}}
\end{table}

\begin{figure}[t!]
\centering
\includegraphics[width=0.70\textwidth]{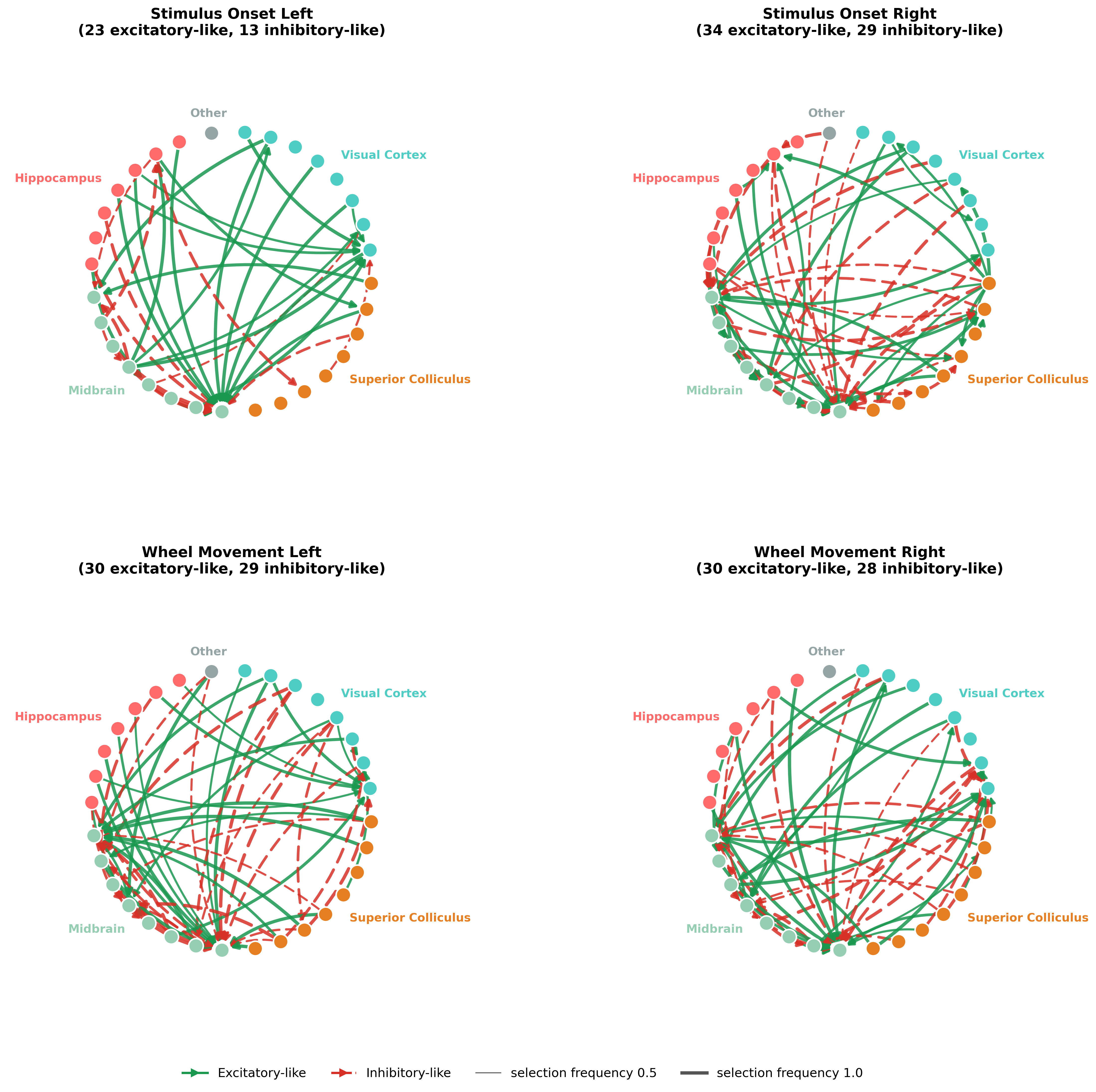}
\caption{Directed networks for left- and right-stimulus trials aligned to stimulus onset (top) and wheel response (bottom). Node colors indicate anatomical groups; solid green and dashed red edges denote excitatory-like and inhibitory-like associations. Full-data edges are retained if selected in at least half of 50 trial subsamples. Edge widths reflect selection frequency.}
\label{fig:networks}
\end{figure}

Table \ref{tab:summary} summarizes the four estimated networks. They contain 36-63 edges, with $50.8\%$-$63.9\%$ classified as excitatory-like, and $68\%$-$75\%$ connecting different anatomical groups. The excitatory-like share is highest for left-stimulus trials aligned to stimulus onset and close to one half in the other conditions.

Figure \ref{fig:networks} reveals a common regional organization across the four conditions. The sampled midbrain neurons are the principal targets, receiving $61\%$-$77\%$ of region-assigned edges. Across conditions, they receive 5-11 edges from visual cortex, 3-12 from superior colliculus, and 7-10 from hippocampal areas, together with 7-14 within-midbrain edges. Return connections from both midbrain and superior colliculus toward visual cortex are also present in every condition. Thus, the networks exhibit substantial convergence toward the sampled midbrain neurons alongside directed associations within and between other regions. This broad organization persists despite substantial variation in individual edges. Six directed neuron-level edges occur in all four networks, and pairwise Jaccard similarities range from $0.15$ to $0.29$, with the greatest similarity between the two wheel-aligned networks. At the region level, seven of the fifteen directed region pairs observed in any condition are shared across all four conditions. These include projections from visual cortex, superior colliculus, and hippocampal areas toward midbrain and return projections from midbrain and superior colliculus toward visual cortex. The regional organization is therefore more consistent across stimulus sides and temporal alignments than individual neuron-level connections.

This analysis illustrates how PPSEM characterizes directed, sign-resolved associations while retaining event-time information. The shared regional organization is broadly consistent with the distributed sensorimotor circuitry described by \citet{steinmetz2019distributed}. However, we remark that these networks come from observational recordings in one session from one animal. Their edges and signs represent model-based conditional dependencies and fitted activity effects, rather than interventionally established causal connections or verified synaptic mechanisms.

\vspace{-0.1in}
\section*{Data Availability Statement}
The data used in this article is publicly available from \url{https://docs.dandiarchive.org/user-guide-using/accessing-data/downloading/}

\bibliographystyle{apa}
\bibliography{ref-ppsem}

\end{document}